\documentclass[prl,twocolumn]{revtex4}
\usepackage{graphicx}
\usepackage{epsfig,color}
\usepackage{amsmath,amssymb,txfonts}

\begin{document}

\title{Unusual Transitions Made Possible by Superoscillations}

\date{\today}

\author{Ran Ber and Moshe Schwartz}
\affiliation{School of Physics and Astronomy, Raymond and Beverly Sackler Faculty of Exact Sciences, Tel Aviv University, Tel-Aviv 69978, Israel.}

\begin{abstract}
We present a physical scenario in which an electromagnetic field in a state consisted of photons whose energies are smaller than the energy gap of a two-state particle, interacts with the particle, for an arbitrarily long duration, as if it was consisted of photons whose energy matches the energy gap of the particle.
This type of interaction is possible when the field is in a specially tailored state whose magnetic field's expectation value is a superoscillatory function.
\end{abstract}

\maketitle

As the time evolution of most quantum systems cannot be calculated analytically, it is very common to use a perturbative approach.
When one wishes to calculate transition probabilities, the mathematical tool usually being used is the time-dependent perturbation theory \cite{Baym1990}.
In many cases, the first order perturbative calculation offers a rather good approximation to the exact result.

When considering a first order solution, and an arbitrary long interaction duration, one might expect that if a certain quantum system is in a state which is a superposition of eigenstates of the unperturbed Hamiltonian, all with energies smaller than the minimal energy gap of another quantum system, the two systems will not interact due to energy conservation.

In this Letter we argue that such interactions are possible.
We demonstrate our argument by describing a quantized electromagnetic (EM) field interacting with a two-state quantum particle.
We set the EM field state to be a superposition of different photon numbers and energies such that all the energies are smaller than the energy gap of a two-state particle.
We then show that, up to first order in perturbation theory, it can interact with the particle as if it was (only) consisted of photons whose energy matches the energy gap of the particle.
This holds for an arbitrary long interaction duration.
This rather strange result is achieved when the expectation value of the magnetic field operator is a superoscillatory function \cite{aharonov1988,Berry1994a}.
This Letter is organized as follows: first, we formulate the EM field's state as a product of coherent states with different frequencies.
Next, we describe the interaction between the field and the two-state particle and show that the transitions described above are possible when the expectation value of the magnetic field operator is a superoscillatory function. 
Finally, we resolve an apparent paradox involving this result and energy conservation.

\emph{Field state as a product of coherent states.}
Consider a quantum EM field. For simplicity, first we assume quasiperiodic boundary conditions in a volume $V$. Later, we would work in the continuum by setting $V\rightarrow \infty$.
Working in the Coulomb gauge, the field operator is
\begin{equation}
\!\mathbf{A}\!\left(\mathbf{r},\! t\right)\!=\!\!\underset{\mathbf{k}p}{\sum}\!\!\sqrt{\frac{2\pi\hbar c^{2}}{V\omega_{k}}}\!\left(a_{\mathbf{k},p}\hat{\mathbf{e}}_{\mathbf{k},p}e^{i\left(\mathbf{k}\cdot\mathbf{r}-\omega_{k}t\right)}\!+\! a_{\mathbf{k},p}^{\dagger}\!\hat{\mathbf{e}}_{\mathbf{k},p}^{*}e^{-i\left(\mathbf{k}\cdot\mathbf{r}-\omega_{k}t\right)}\right)\!,
\end{equation}
where $a_{\mathbf{k},p}$ and $a_{\mathbf{k},p}^{\dagger}$ are annihilation and creation operators of photons with momentum $\mathbf{k}$ and polarization $p$ respectively, $\hat{\mathbf{e}}_{\mathbf{k},p}$ is a polarization vector and $c$ is the speed of light.
The electric and magnetic fields are then $\mathbf{E}\left(\mathbf{r},t\right)=-\frac{1}{c}\frac{\partial\mathbf{A}\left(\mathbf{r},t\right)}{\partial t}$ and $\mathbf{B}\left(\mathbf{r},t\right)=\mathbf{\nabla}\times\mathbf{A}\left(\mathbf{r},t\right)$.
We wish to find a field state which satisfies 
\begin{equation}
\langle B_{x}\left(\mathbf{r},t\right)\rangle=F\left(z-ct\right),\label{Bx=F(z-ct)}
\end{equation}
for an arbitrary $F\left(z-ct\right)$.
It is well known that coherent states of the form 
\begin{equation}
\left|\alpha_{\mathbf{k},p}\right\rangle =e^{-\frac{1}{2}\left|\alpha\right|^{2}}\underset{n}{\sum}\frac{\alpha^{n}}{\sqrt{n!}}\left|n_{\mathbf{k},p}\right\rangle,
\end{equation}
where $\left|n_{\mathbf{k},p}\right\rangle $ is a field state consisted of $n$ photons with momentum $\hbar\mathbf{k}$ and polarization $p$, are states whose expectation values are harmonic waves.
In order to demonstrate this property explicitly, WLOG, we set the momentum and polarization such that $\hat{\mathbf{k}}=\hat{\mathbf{z}}$ and $\hat{\mathbf{p}}=\hat{\mathbf{y}}$.
Then, using
$a_{\mathbf{k}',p'}\left|\alpha_{\mathbf{k},p}\right\rangle =\delta_{\mathbf{k},\mathbf{k}'}\delta_{p,p'}\alpha_{\mathbf{k},p}\left|\alpha_{\mathbf{k},p}\right\rangle $
we obtain 
\begin{eqnarray}
 &  & \!\left\langle \alpha_{k,p}\right\vert B_{x}\left(\mathbf{r},t\right)\left\vert \alpha_{k,p}\right\rangle =\nonumber \\
 &  & \!\sqrt{\frac{8\pi\hbar\omega_{k}}{V}}\!\left(\text{Re}\!\left[\alpha_{k}\right]\!\sin\!\left(kz\!-\!\omega_{k}t\right)\!+\!\text{Im}\!\left[\alpha_{k}\right]\!\cos\!\left(kz\!-\!\omega_{k}t\right)\right)\!,\label{Ax}
\end{eqnarray}
as required.
Next, in order to obtain a state which satisfies Eq. (\ref{Bx=F(z-ct)}) we rewrite $F\left(z-ct\right)$ as a Fourier series: 
\begin{equation}
F\!\left(z-ct\right)=\underset{k}{\sum}\!\left(A_{n}\cos\!\left(k_{n}z-\omega_{k_{n}}t\right)+B_{n}\sin\!\left(k_{n}z-\omega_{k_{n}}t\right)\right)\!,\label{F(z)}
\end{equation}
where $A_{n}=\frac{2}{V^{1/3}}\int dzF\left(z\right)\cos\left(k_{n}z\right)$ and $B_{n}=\frac{2}{V^{1/3}}\int dzF\left(z\right)\sin\left(k_{n}z\right)$ (the notation $F\left(z,0\right)=F\left(z\right)$ is used).
Comparing $A_{n}$ and $B_{n}$ to the corresponding terms in Eq. (\ref{Ax}) leads to 
\begin{equation}
\left|\alpha\right\rangle = \underset{k>0}{\Pi}\left|i\sqrt{\frac{V^{1/3}}{2\pi\hbar\omega_{k}}}\tilde{F}\left(k\right)\right\rangle,
\end{equation}
where $\tilde{F}\left(k\right)\equiv\int dzF\left(z\right)e^{-ikz}$.
This state is a product of coherent states, $\left|\alpha_{k}\right\rangle $, each with a different frequency $\omega_{k}$.
When $\alpha_{k}\gg1\forall k$ this can be regarded as a ``classical'' state.

\emph{The interaction.}
We wish to couple the EM field to a quantum particle whose free Hamiltonian is $H_{\text{par}}=H_{\text{s}}\left(\mathbf{r}_{\text{m}},\mathbf{p}_{\text{m}}\right)+\frac{1}{2}\hbar\Omega\sigma_{z}$, where $H_{\text{s}}$ is the spatial term of the Hamiltonian, $\mathbf{r}_{\text{m}}$, $\mathbf{p}_{\text{m}}$ and $\mathbf{\sigma}$ are the particle's position, momentum and spin operators respectively, and $\Omega$ is the particle's frequency.
We use the standard field-matter interaction which is
\begin{eqnarray}
H_{\text{int}} & = & \!\!\int\! d^{3}\mathbf{r}\left(-\frac{e}{c}\mathbf{A}\left(\mathbf{r}\right)\mathbf{j}\left(\mathbf{r}\right)+\frac{e^{2}}{2mc^{2}}\mathbf{A}^{2}\left(\mathbf{r}\right)\rho\left(\mathbf{r}\right)\right)\nonumber \\
 &  & -\frac{e\hbar}{2mc}\mathbf{B}\left(\mathbf{r}_{\text{m}}\right)\cdot\mathbf{\sigma},
\end{eqnarray}
where $m$ is the particle's mass, $\rho\left(\mathbf{r}\right)\equiv\delta\left(\mathbf{r}-\mathbf{r}_{\text{m}}\right)$ is the particle's density and $\mathbf{j}\left(\mathbf{r}\right)\equiv\frac{1}{2m}\left(\mathbf{p}{}_{\text{m}}\delta\left(\mathbf{r}-\mathbf{r}{}_{\text{m}}\right)+\delta\left(\mathbf{r}-\mathbf{r}{}_{\text{m}}\right)\mathbf{p}{}_{\text{m}}\right)$ is the particle's currant.
Since we are only interested in the field-spin interaction, we shall consider only the magnetic interaction from now on. For simplicity, we also set $H_\text{s}=0$. 
Assuming the interaction is turned on sharply at $t=0$, but is independent of time afterwards, we get in the interaction picture
\begin{equation}
H_{I}=-\frac{e\hbar}{2mc}\mathbf{B}\left(\mathbf{r}_{\text{m}},t\right)\cdot\mathbf{\sigma}\left(t\right).
\end{equation}
Next, we assume that the field is initially in state $\left|\alpha\right\rangle $, and that the particle is initially in state $\left|M_{i}\right\rangle =\left|\Psi,\downarrow\right\rangle $, where $\left|\Psi\right\rangle$ is the spatial component of the state and $\left|\downarrow\right\rangle$ denotes the ground state of the spin. 
The interaction then leads (to first order in perturbation theory) to the state 
\begin{equation}
\left|\text{EM},\sigma\right\rangle _{t}=\left|\alpha,M_{i}\right\rangle \!+\!\frac{ie}{2mc}\!\int_{0}^{t}\! dt'B_{x}\left(\mathbf{r}_{\text{m}},t'\right)\sigma_{x}\left(t'\right)\left|\alpha,M_{i}\right\rangle .
\end{equation}
The probability to measure the particle in state $\left|M_{f}\right\rangle =\left|\Psi,\uparrow\right\rangle$ is therefore 
\begin{eqnarray}
P\left(t\right) & = & \left(\frac{e}{2mc}\right)^{2}\!\sum_{\xi}\! \left|\left\langle M_{f},\xi\right|\!\int_{0}^{t}\! dt'B_{x}\!\left(\mathbf{r}_{\text{m}},t'\right)\sigma_{x}\!\left(t'\right)\left|M_{i},\alpha\right\rangle \right|^{2}\nonumber \\
 & = & \left(\frac{e}{2mc}\right)^{2}\!\!\int_{0}^{t}\! dt'\!\int_{0}^{t}\! dt''e^{i\Omega\left(t'-t''\right)}\nonumber \\
 &  & \times\left\langle \Psi,\alpha\right|B_{x}\!\left(\mathbf{r}_{\text{m}},t'\right)\left\vert \Psi \right\rangle \left\langle \Psi \right\vert B_{x}\!\left(\mathbf{r}_{\text{m}},t''\right)\left|\Psi,\alpha\right\rangle, \label{P(t)(1)}
\end{eqnarray}
where $\left\lbrace \left\vert \xi \right\rangle\right\rbrace$ is an orthonormal basis. Assuming $\left|\Psi\right\rangle $ is very localized around $z_{0}$,
i.e., $\Delta z_{\text{m}}^{2}\ll k_{0}^{-2}$ we obtain 
\begin{equation}
\left\langle \Psi\right|e^{ikz_{\text{m}}}\left|\Psi\right\rangle \cong e^{ikz_0},\label{exp(ikr)}
\end{equation}
where $z_0=\left\langle \Psi\right|z_{\text{m}}\left|\Psi\right\rangle$. Therefore
\begin{equation}
P\!\left(t\right) \!=\! \left(\frac{e}{2mc}\right)^{2}\!\!\!\int_{0}^{t}\!\!\! dt'\!\!\!\int_{0}^{t}\!\!\! dt''\!e^{i\Omega\left(t'-t''\right)}\!\left\langle\alpha\right|\!B_{x}\!\left(z_0,\!t'\right)\!B_{x}\!\left(z_0,\!t''\right)\!\left|\alpha\right\rangle\!.\label{P(t)(2)}
\end{equation}
For simplicity, from now on we shall assume a large volume.
In this limit $\sum_{k}\rightarrow\frac{1}{2\pi}\int V^{1/3}dk$ and $\left[a_{k},a_{k'}^{\dagger}\right]\rightarrow2\pi V^{-1/3}\delta_{k-k'}$.
Plugging the latter commutation relation in the spectral representation of $\mathbf{B}$ we obtain 
\begin{eqnarray}
 &  & \left\langle \alpha\right|B_{x}\left(z_0,t'\right)B_{x}\left(z_0,t''\right)\left|\alpha\right\rangle \cong\nonumber \\
 &  & F\!\left(z_{0}\!-\! ct'\right)\! F\!\left(z_{0}\!-\! ct''\right)\!+\!\frac{\hbar}{V^{2/3}c}\!\!\int\!\!\! d\omega\omega e^{i\omega\left(t''-t'\right)}.
\end{eqnarray}
The second term could be zeroed in the limit of a large volume.
Plugging this result back in Eq. (\ref{P(t)(2)}) leads to
\begin{equation}
P\left(t\right)=\left(\frac{e}{2mc}\right)^{2}\left|\int_{0}^{t}dt'F\left(z_{0}-ct'\right)e^{i\Omega t'}\right|^{2}.\label{P(t)(3)}
\end{equation}
Next, we consider a field state consisted of photons all with momentum $k\in\left[0,k_{0}\right]$. This amounts to having $F\left(z\right)$ such that $\tilde{F}\left(k\right)$ only has support in $k\in\left[0,k_{0}\right]$.
Under usual circumstances, it is clear that a wave packet corresponding to such $F\left(z\right)$ will only interact with particles whose energy gap is $E\in\left[0,\hbar ck_{0}\right]$.
However, we shall now show that using a superoscillatory $F\left(z\right)$ one could interact such a wave packet with particles whose energy gap is $\hbar\Omega\equiv E>\hbar ck_{0}$.

\emph{Superoscillations.}
Superoscillatory functions are functions which oscillate faster than their fastest Fourier component.
These functions can superoscillate in an arbitrarily large, but not infinite domain.
Since superoscillations are due to a destructive interference, outside the superoscillatory domain these functions exhibit exponentially larger amplitudes.
Usually these larger amplitudes are present both before and after the superoscillatory domain.
However, some superoscillatory functions have an exponentially larger amplitude only after (or before) the superoscillatory domain.
In this work, we shall use such functions.

Consider the following function \cite{Berry1994,Reznik1997,Ber2014,Ber2014a}: 
\begin{equation}
F\left(z\right)=\frac{D}{2\delta\sqrt{2\pi}}\int_{0}^{2\pi}d\alpha e^{izk_{0}\left(\frac{1+\cos\alpha}{2}\right)}e^{-\frac{i}{\delta^{2}}\cos\left(\alpha-iA\right)},\label{F(z)so}
\end{equation}
where $D$, $\delta$ and $A$ are some constants and $k=k_{0}\left(1+\cos\alpha\right)/2$.
While $\tilde{F}\left(k\right)$ has support only in $\left[0,k_{0}\right]$, we shall now prove that $F\left(z\right)$ oscillates arbitrarily fast.
Performing the integration explicitly we obtain
\begin{eqnarray}
F\left(z\right) & = & \frac{D\sqrt{\pi}}{\sqrt{2}\delta}e^{\frac{1}{2}izk_{0}}\nonumber \\
 &  & \times J_{0}\left(\frac{1}{\delta^{2}}\sqrt{1\!-\!\delta^{2}zk_{0}\cosh\left[A\right]\!+\!\frac{1}{4}\delta^{4}z^{2}k_{0}^{2}}\right),
\end{eqnarray}
where $J_0$ is the zeroth Bessel function. For $z<0$, using the asymptotic form of the Bessel function for $\delta\ll1$ we get
\begin{widetext}
\begin{equation}
F\left(z\right)\cong\frac{D}{\left(1-\delta^{2}zk_{0}\cosh\left[A\right]+\frac{1}{4}\delta^{4}z^{2}k_{0}^{2}\right)^{\frac{1}{4}}}e^{\frac{1}{2}izk_{0}}\cos\left(\frac{1}{\delta^{2}}\sqrt{1-\delta^{2}zk_{0}\cosh\left[A\right]+\frac{1}{4}\delta^{4}z^{2}k_{0}^{2}}-\frac{\pi}{4}\right).
\end{equation}
\end{widetext}
We wish this function to superoscillate in $z\in\left[-z_c,0\right]$. To this end we take $\delta^{2}\ll\frac{1}{z_c k_0 \cosh\left[A\right]}$.
Then, in this domain, the function reduces to 
\begin{equation}
F\left(z\right)\cong De^{\frac{1}{2}izk_{0}}\cos\left(\frac{1}{\delta^{2}}-\frac{1}{2}zk_{0}\cosh\left[A\right]-\frac{\pi}{4}\right).
\end{equation}
Redefining $F\left(z\right)$ to be the summation of two such functions, one having $\delta^{-2}=2\pi m+\pi/4$ and the other $D\rightarrow\pm iD$ and $\delta^{-2}=2\pi m+3\pi/4$, where $m\gg1$ we obtain 
\begin{equation}
F\left(z\right)\cong De^{\frac{1}{2}izk_{0}\left(1\pm\cosh\left[A\right]\right)}.
\end{equation}
This function oscillates at frequency $k'=\frac{1}{2}k_{0}\left(1+\cosh\left[A\right]\right)$.
By increasing $A$, one can set these superoscillations to be arbitrarily fast.
The superoscillatory domain is finite, however one could set it to be arbitrarily large by decreasing $\delta$.

Superoscillations come at the price of an exponential growth outside the superoscillatory domain, in our case with maximum $\sim\frac{D}{2\sqrt{\sinh[A]}}\exp\left(\frac{\sinh[A]}{\delta^{2}}\right)$ occurring at $z=\frac{2\cosh[A]}{k_{0}\delta^{2}}>0$.
Bellow the superoscillatory domain, as well as above the exponential growth, the function gradually obtains regular (slower) oscillations.
In the limit $\left|z\right|\gg4\cosh\left[A\right]/(\delta^{2}k_{0})$, it behaves like $\frac{1}{\sqrt{z}}\sin\left(zk_{0}\right)$.
This means that the proposed superoscillatory function $F\left(z\right)$ is not normalizable.
In order to make it normalizable, one may convolute $\tilde{F}\left(k\right)$ with a smooth function $\tilde{h}\left(k\right)$ which has a very small support around $k=0$.
This amounts to replacing $F\left(z\right)$ by $F\left(z\right)h\left(z\right)$. Since $\tilde{h}\left(k\right)$ is smooth, the new function decreases fast enough as $z\rightarrow\pm\infty$ to be normalizable.

\begin{figure}
\includegraphics[scale=0.225]{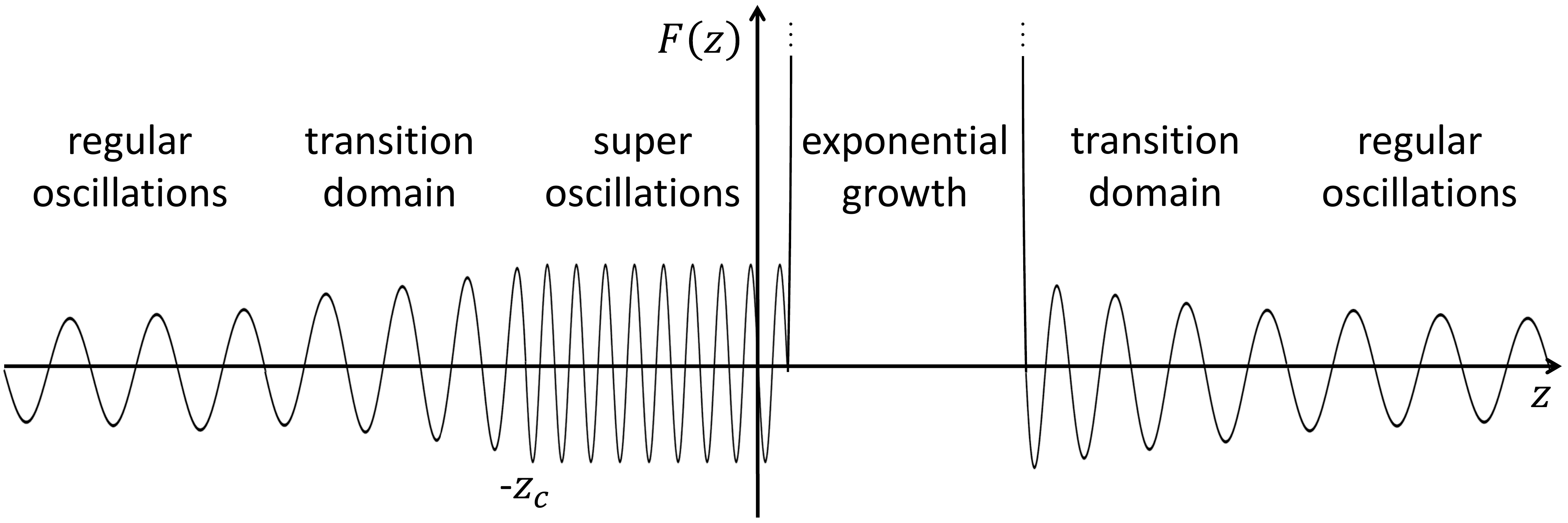}
\caption{A schematic plot of the superoscillatory function that we use: the function obtains its exponential growth at $z>0$. Since the wave is moving to the right, it does not interact with the particle there. At $-z_c<z<0$ the function superoscillates, and as $z\rightarrow\pm\infty$ it gradually obtains regular (slower) oscillations and decays.}
\label{superoscillatory function}
\end{figure}

\emph{Transition probability.}
The resulting function (plotted in Fig. (\ref{superoscillatory function})) is band limited to $\left[0,k_{0}\right]$, however it can oscillate in an arbitrarily large domain $\left[-z_c,0\right]$ arbitrarily fast, and it is normalizable.
As the EM wave is moving to the right, placing the detector at $z_0=0$ guaranties that it will first interact with the EM wave where the wave superoscillates and then where it gradually gains its regular oscillations. It will not interact with the wave where it grows exponentially.

Substituting in Eq. (\ref{P(t)(3)}) $F\left(z\right)\propto\sin\left(\frac{\Omega}{c}z\right)$ for $\left[-z_{c},0\right]$, then for $t\le z_c/c$ we finally obtain 
\begin{equation}
P\left(t\right)\propto\left|\int_{0}^{t}dt'\left(e^{-i\Omega t'}- e^{i\Omega t'}\right)e^{i\Omega t'}\right|^{2}\propto t^2,
\end{equation}
i.e., for as long as the EM field state is superoscillatory, it interacts with the particle exactly as if it was a monochromatic EM field state of frequency $\Omega$: it only interacts with particles whose energy gap is $E=\hbar\Omega$, and it excites the particles at the same rate as a monochromatic field state of the same frequency would.
Since $F\left(z\right)$ decays fast below the superoscillatory domain, the perturbation can be small even for $t\rightarrow\infty$, keeping the first order solution viable.

\emph{Explicit calculation of the EM field's energy.}
Since the EM field and the spin are coupled for an arbitrarily long duration, the energy absorbed by the spin could not originate from the activation/shutting of the coupling.
We therefore expect the energy absorbed by the particle to come from the EM field.
In order to verify this assumption, we shall now show explicitly that once the particle is measured in the excited state, the EM field loses energy that matches the energy gap of the particle.
First, we calculate the EM field's energy before the interaction.
Using $a_{k}\left|\alpha\right\rangle =i\!\sqrt{\frac{V^{1/3}}{2\pi\hbar\omega_{k}}}\tilde{F}\left(k\right)\left|\alpha\right\rangle $  we get
\begin{equation}
\left\langle E_{b}\right\rangle =\frac{V^{2/3}}{4\pi^{2}}\int dk\left|\tilde{F}\left(k\right)\right|^{2}.
\end{equation}
Once the spin is measured in the $\left|\uparrow\right\rangle $ state, the EM field's state becomes
\begin{equation}
\left|\text{EM}_{a}\right\rangle =\frac{\int_{0}^{t}\! dt'B_{x}\left(\mathbf{r}_{\text{m}},t'\right)\sigma_{x}\left(t'\right)\left|\alpha,M_{i}\right\rangle }{\sqrt{\left\langle \alpha,M_{i}\right|\left|\int_{0}^{t}\! dt'B_{x}\left(\mathbf{r}_{\text{m}},t'\right)\sigma_{x}\left(t'\right)\right|^{2}\left|\alpha,M_{i}\right\rangle }},
\end{equation}
thus its energy is
\begin{align}
\left\langle E_{a}\right\rangle  & =\frac{V^{1/3}\!\int\! dk\hbar\omega_{k}}{2\pi\left\langle \Psi,\alpha\right|\left|\int_{0}^{t}\! dt'B_{x}\left(\mathbf{r}_{\text{m}},t'\right)e^{i\Omega t'}\right|^{2}\left|\Psi,\alpha\right\rangle }\nonumber \\
 & \!\times\!\left\langle \Psi\!,\!\alpha\right|\!\!\!\int_{0}^{t}\!\!\! dt'B_{x}\!\left(\mathbf{r}_{\text{m}},t'\right)e^{-i\Omega t'}\!a_{k}^{\dagger}a_{k}\!\!\!\int_{0}^{t}\!\!\! dt''B_{x}\!\left(\mathbf{r}_{\text{m}},t''\right)e^{i\Omega t''}\left|\Psi\!,\!\alpha\right\rangle \!.\nonumber \\
\end{align}

Next, using $\left[a_{k},B_{x}\left(\mathbf{r}_{\text{m}},t\right)\right]=i\sqrt{\frac{2\pi\hbar\omega_{k}}{V}}e^{-i\left(kz_{\text{m}}-\omega_{k}t\right)}$ we get
\begin{widetext}
\begin{eqnarray}
\left\langle E_{a}\right\rangle  & = & \frac{V^{1/3}\hbar\int dk\omega_{k}\left\langle \Psi,\alpha\right|a_{k}^{\dagger}\left|\int_{0}^{t}dt'B_{x}\left(\mathbf{r}_{\text{m}},t'\right)e^{i\Omega t'}\right|^{2}a_{k}\left|\Psi,\alpha\right\rangle }{2\pi\left\langle \Psi,\alpha\right|\left|\int_{0}^{t}\! dt'B_{x}\left(\mathbf{r}_{\text{m}},t'\right)e^{i\Omega t'}\right|^{2}\left|\Psi,\alpha\right\rangle }\nonumber \\
 &  & +\frac{i\sqrt{\frac{\hbar^{3}}{2\pi V^{1/3}}}\int dk\omega_{k}^{3/2}\int_{0}^{t}dt'\int_{0}^{t}dt''e^{i\Omega\left(t''-t'\right)}\left\langle \Psi,\alpha\right|\left(e^{-i\left(kz_{\text{m}}-\omega_{k}t''\right)}a_{k}^{\dagger}B_{x}\left(\mathbf{r}_{\text{m}},t'\right)-e^{i\left(kz_{\text{m}}-\omega_{k}t'\right)}B_{x}\left(\mathbf{r}_{\text{m}},t''\right)a_{k}\right)\left|\Psi,\alpha\right\rangle }{\left\langle \Psi,\alpha\right|\left|\int_{0}^{t}\! dt'B_{x}\left(\mathbf{r}_{\text{m}},t'\right)e^{i\Omega t'}\right|^{2}\left|\Psi,\alpha\right\rangle }\nonumber \\
 &  & +\frac{\hbar^{2}\int dk\omega_{k}^{2}\int_{0}^{t}dt'\int_{0}^{t}dt''e^{i\left(\Omega+\omega_{k}\right)\left(t''-t'\right)}}{V^{2/3}\left\langle \Psi,\alpha\right|\left|\int_{0}^{t}\! dt'B_{x}\left(\mathbf{r}_{\text{m}},t'\right)e^{i\Omega t'}\right|^{2}\left|\Psi,\alpha\right\rangle }.
\end{eqnarray}
\end{widetext}
We proceed by defining $\left\langle E_a \right\rangle\equiv I_1+I_2+I_3$ and calculating the $3$ terms.
For the first term, again, using $a_{k}\left|\alpha\right\rangle =i\!\sqrt{\frac{V^{1/3}}{2\pi\hbar\omega_{k}}}\tilde{F}\left(k\right)\left|\alpha\right\rangle $, we obtain
\begin{equation}
I_{1}=\left\langle E_{b}\right\rangle.
\end{equation}
For the second term a straight forward calculation yields
\begin{align}
I_{2} & =\frac{\hbar c\int_{0}^{t}dt'\int_{0}^{t}dt''}{\left\langle \Psi\!,\!\alpha\right|\left|\int_{0}^{t}\! dt'B_{x}\left(\mathbf{r}_{\text{m}},t'\right)e^{i\Omega t'}\right|^{2}\left|\Psi\!,\!\alpha\right\rangle }\nonumber \\
 & \!\times\left\langle \Psi\!,\!\alpha\right|F'\!\left(z_{\text{m}}\!-\! ct'\right)\! B_{x}\!\left(\mathbf{r}_{\text{m}},t''\right)\sin\!\left(\Omega\left(t''\!-\! t'\right)\right)\left|\Psi\!,\!\alpha\right\rangle.
\end{align}
Substituting $\left\langle \Psi,\alpha\right|B_{x}\left(\mathbf{r}_{\text{m}},t\right)\left|\Psi,\alpha\right\rangle \cong F\left(-ct\right)$,
$F\left(z\right)\propto\sin\left(\frac{\Omega}{c}z\right)$
and $t\gg\frac{2\pi}{\Omega}$ we get 
\begin{eqnarray}
I_{2} & = & \frac{-E\!\int_{0}^{t}\! dt'\!\int_{0}^{t}\! dt''\cos\left(\Omega t'\right)\sin\left(\Omega t''\right)\sin\left(\Omega\left(t''\!-\! t'\right)\right)}{\left\vert \int_{0}^{t}dt'\sin\left(\Omega t'\right)e^{-i\Omega t'}\right\vert ^{2}}\nonumber \\
 & \rightarrow & -E.
\end{eqnarray}
For the third term, a similar substitution yields $I_{3}\rightarrow0$.
Therefore we finally get
\begin{equation}
\left\langle E_{a}\right\rangle =\left\langle E_{b}\right\rangle -E,
\end{equation}
as expected.

In this Letter we have studied an interaction between an EM field and a quantum particle.
we have assumed a first order perturbative solution and allowed an arbitrarily long interaction duration, yet still we have described a scenario in which an EM field in a state consisted of photons, all with energy (arbitrarily) smaller than the particle's energy gap, excite the particle as if the energy of the photons matched the particle's energy gap.
This unusual result was made possible due to a special field state for which the magnetic field operator's expectation value was a superoscillatory function. 

This result seems to contradict energy conservation since the mean energy of photons, all with energy smaller than the energy gap of the two-state particle is necessarily smaller than the energy gap of the particle.
This apparent paradox is resolved by noting that the common interpretation of ``absorption'' and ``emission'' is over-simplified when the process involves transitions between superposition of eigenstates of the unperturbed Hamiltonian, rather than a single eigenstate.
This is because the creation and annihilation operators, $a^{\dagger}$ and $a$, change the weights of the different components in the superposition in addition to the creation and annihilation of photons.
The superoscillations are used to ``fine-tune'' the modification of the weights which leads to this dramatic effect.

It is interesting to note that once the particle gets excited it can spontaneously decay to the ground state by emitting a photon, whose energy matches the energy gap of the particle.
This can be regarded as a mechanism that transforms a ``superoscillatory photon'' into a ``real photon'' of the same frequency.

\emph{Acknowledgements.}
RB acknowledges the support of the Israel Science Foundation Grant No. 1311/14.

\bibliography{ref}

\end{document}